\documentclass[twocolumn,prb,superscriptaddress]{revtex4-2}
\newcommand{\pagenumbaa}{1}
\usepackage{lipsum}	
\usepackage[english]{babel}
\usepackage{graphicx}
\usepackage{framed}
\usepackage{amsmath}
\usepackage{amsthm}
\usepackage{amssymb}
\usepackage{amsfonts}
\usepackage{enumerate}
\usepackage[utf8]{inputenc}
\usepackage{datetime}
\usepackage{indentfirst}
\usepackage{float}
\usepackage{physics}
\usepackage{siunitx}
\usepackage{adjustbox}
\usepackage{colortbl}
\usepackage[table,dvipsnames]{xcolor}
\usepackage[colorlinks,citecolor=blue,linkcolor=blue,urlcolor=blue]{hyperref}
\usepackage[capitalize]{cleveref}

\begin{document}

\title{Gate Tunable Josephson Diode in Proximitized InAs Supercurrent Interferometers}

\author{Carlo Ciaccia}
\email[E-mail: ]{Carlo.Ciaccia@unibas.ch}
\affiliation
{Quantum- and Nanoelectronics Lab, Department of Physics, University of Basel, 4056 Basel, Switzerland}

\author{Roy Haller}
\affiliation
{Quantum- and Nanoelectronics Lab, Department of Physics, University of Basel, 4056 Basel, Switzerland}

\author{Asbj\o{}rn C. C. Drachmann}
\affiliation
{Center for Quantum Devices, Niels Bohr Institute, University of Copenhagen, 2100 Copenhagen, Denmark}
\affiliation
{NNF Quantum Computing Programme, Niels Bohr Institute, University of Copenhagen, 2100 Copenhagen, Denmark}

\author{Tyler Lindemann}
\affiliation
{Department of Physics and Astronomy, Purdue University, West Lafayette, Indiana 47907, USA}
\affiliation
{Birck Nanotechnology Center, Purdue University, West Lafayette, Indiana 47907, USA}

\author{Michael J. Manfra}
\affiliation
{Department of Physics and Astronomy, Purdue University, West Lafayette, Indiana 47907, USA}
\affiliation
{Birck Nanotechnology Center, Purdue University, West Lafayette, Indiana 47907, USA}
\affiliation
{Elmore Family School of Electrical and Computer Engineering, Purdue University, West Lafayette, Indiana 47907, USA}
\affiliation
{School of Materials Engineering, Purdue University, West Lafayette, Indiana 47907, USA}

\author{Constantin Schrade}
\affiliation
{Center for Quantum Devices, Niels Bohr Institute, University of Copenhagen, 2100 Copenhagen, Denmark}

\author{Christian Sch{\"o}nenberger}
\email[E-mail: ]{Christian.Schoenenberger@unibas.ch}
\affiliation
{Quantum- and Nanoelectronics Lab, Department of Physics, University of Basel, 4056 Basel, Switzerland}
\affiliation
{Swiss Nanoscience Institute, University of Basel, 4056 Basel, Switzerland}

\begin{abstract}
The Josephson diode (JD) is a non-reciprocal circuit element that supports a larger critical current in one direction compared to the other. This effect has gained a growing interest because of promising applications in super\-conducting electronic circuits with low power consumption. Some implementations of a JD rely on breaking the inversion symmetry in the material used to realize Josephson junctions (JJs), but recent theoretical proposals have suggested that the effect can also be engineered by combining two JJs hosting highly transmitting Andreev bound states in a Superconducting Quantum Interference Device (SQUID) at a small, but finite flux bias.
We have realized a SQUID with two JJs fabricated in a proximitized InAs two-dimensional electron gas (2DEG). We demonstrate gate control of the diode efficiency from zero up to around $30$\% at specific flux bias values which comes close to the maximum of $\sim 40$\% predicated in Ref. [R.~S.\ Souto, M. Leijnse and C. Schrade, Phys. Rev. Lett. {\bf 129}, 267702 (2022)]. The key ingredients to the JD effect in the SQUID arrangement is the presence of highly transmitting channels in the JJs, a flux bias and an asymmetry between the two SQUID arms.
\end{abstract}

\maketitle

\setcounter{page}{\pagenumbaa}
\thispagestyle{plain}


\section{Introduction}
\label{sec:Introduction}
\vspace{-0.3cm}

A widely used device in semiconductor electronics is the $p-n$ junction, which is a nonreciprocal element with regards to current flow, able to conduct current primarily in one direction. The presently ongoing rapid scaling of quantum computers will require low-dissipative control electronics that operate close to the quantum chip at low temperatures. These requirements have renewed the question whether there exists a super\-conducting equivalent of the diode, namely a device that supports a larger super\-current in one direction than in another: the Josephson Diode (JD)~\cite{Kokkeler2022,Zhang2022}.

In a conventional Josephson Junction (JJ)~\cite{Josephson1962}, the current-phase relation (CPR) is sinusoidal $I = I_{\rm c}\sin(\varphi)$, with $I_{\rm c}$ being the critical current of the junction and with the ground state corresponding to zero phase bias $\varphi_0=0$. For this conventional case, the positive critical current, $I_{\textrm{c}}^{+}=\mathrm{max}_{\varphi}[I(\varphi)]$ is obviously equal to the negative one $I_{\textrm{c}}^{-}=\left|\mathrm{min}_{\varphi}[I(\varphi)]\right|$. Since the critical super\-current is reciprocal, there is no super\-conducting diode-effect (SDE).

A general CPR can have a more complex dependence on the phase~\cite{Golubov2004}. But in general, $I(\varphi)$ is a $2\pi$-periodic function and if either time-reversal symmetry or inversion symmetry is preserved, it is an odd function: $I(-\varphi)=-I(\varphi)$~\cite{Kokkeler2022}. It can therefore be written as a Fourier series composed of $sin(k\varphi)$ terms where $k$ is a positive integer and the terms for $k>1$ are higher harmonics. If higher harmonics are present, the CPR is called non-sinusoidal~\cite{Koops1996,Rocca2007}. Such a CPR still does not display a SDE.

A necessary but not sufficient condition for the SDE to occur is that time-reversal symmetry is broken. This can be achieved either by an external magnetic-field of or by means of ferromagnetic elements built into the device.
%
S-F-S junctions, where F (S) denotes a ferromagnet (super\-conductor) were proposed~\cite{Buzdin1982,Buzdin2003}, and  experimentally studied in various configurations~\cite{Ryazanov2001,Kontos2002,Gingrich2016}. These junctions typically display a $\pi$ shift in the CPR and are thus know as $\pi$-junctions. The energy ground state moves from $\varphi_0 = 0$ to $\varphi_0 = \pi$. Despite the presence of a magnetic field and time-reversal symmetry thus being broken, these junctions do not display a SDE.

Both inversion symmetry and time-reversal symmetry are broken in so-called anomalous JJs, also known as $\varphi_0$ junctions, where the ground state of the junction has an `anomalous' shift to $\varphi_0$ with $0<\varphi_0<\pi$~\cite{Sickinger2012}. This situation is achieved in multiband conductors with spin-orbit interaction~\cite{Krive2004,Reynoso2008,Buzdin2008,Zazunov2009,Yokoyama2013,Wakatsuki2017,Turini2022}. Evidence for $\varphi_0$ junctions has been found in experiments with nanowires with strong spin-orbit interaction~\cite{Szombati2016} and in planar Josephson junction arrays~\cite{Baumgartner2022}. An anomalous JJ is also a necessary condition, but on its own not sufficient.
Indeed, a CPR of the form $I(\varphi)=I_c sin(\varphi - \varphi_0)$ with $0<\varphi_0<\pi$ is an anomalous JJ, but still with $I_{\rm c}^{+}=I_{\rm c}^{-}$.

The SDE has been observed in materials that display magneto-chiral anisotropy. Here, the normal-state resistivity itself depends on the sign of the current density and the sign of the magnetic field~\cite{Rikken2001,Rikken2005,Wakatsuki2017}. While this is a small effect in normal metals, it can become large at the transition to a super\-conducting state~\cite{Ando2020,Daido2022,Baumgartner2022b}. Recently, a large SDE was also observed in a 2D NbSe$_2$ super\-conductor with applied out-of-plane magnetic field~\cite{Bauriedl2022} and even in field-free situations~\cite{Narita2022,Jeon2022,Kokkeler2022} including twisted graphene~\cite{Wu2022,Lin2022,Diez-Merida2023}.

Further studies have also considered, among others, polarized super\-currents, magnetic domain walls, vortex pinning, combination of s-wave and p-wave pairing, as well as finite-momentum pairing as the origin of a SDE~\cite{Yuan2022,Davydova2022,Suri2022,Pal2022}. A SDE was even reported in a scanning-probe microscopy study where a single magnetic impurity was addressed on the surface of a super\-conductor~\cite{Trahms2022}.

Lastly, topological materials with helical edge states can carry super\-currents with a strong SDE~\cite{Chen2018,Kononov2020,Legg2022,Cuozzo2023}. This is evidenced in the highly asymmetric Fraunhofer pattern with the property that $I_{\textrm{c}}(B) \neq I_{\textrm{c}}(-B)$, where $B$ is the magnetic field.
This arises because of lack of inversion symmetry between the super\-current flowing along the two edges of the crystal~\cite{Chen2018}. This situation is very much alike an asymmetric SQUID.

Already in the 1970s, when super\-conducting interference devices were studied in great detail using tunnel junctions, point contact structures and Dayem bridges, it was recognized that the critical current of a SQUID can become non-reciprocal~\cite{Fulton1970,Fulton1972,Tsang1975,Barone1982}.
The origin was understood to emerge from an asymmetry in the two SQUID arms, but the arms needed to have a non-negligible loop inductance, too. Although the CPR of each single junction was sinusoidal, the CPR became non-reciprocal for the SQUID device due to asymmetric loop inductances.

Today, tunable super\-conductor-semiconductor hybrid devices have become a flourishing research topic~\cite{Buitelaar2002,Doh2005,Larsen2015,Prada2020,Burkard2020}.
In particular, in JJ made of semiconducting weak links, the magnitude of the super\-current is tunable by local gate electrodes and, in some devices, the shape of the CPR can be tuned  from sinusoidal to highly non-sinusoidal. Consequently, these devices provide a platform for the engineering of the SDE with unprecedented tunability. This has recently been investigated theoretically in Ref.~\cite{Souto2022,Fominov2022}. It has been shown that one can achieve a large SDE by combining two non-sinusoidal JJs in a dc-SQUID at finite flux bias even with negligible loop inductances. In this case, the non-reciprocal transport $I_c^+ ¨\not= I_c^-$ originates from the interference between higher-order harmonics in CPR of the JJs.


In the current work, we use gate-controlled JJs fabricated in an InAs 2DEG proximitized by an Al layer~\cite{Lee2019,Nichele2020}. These rather wide junctions contain many channels with a distribution of transmission eigenvalues. The non-sinusoidal character is due to highly transmissive channels that are present in these devices~\cite{Dorokhov1984,Nanda2017,Bretheau2017,Manjarres2020,Indolese2020,Haller2022}. By tuning the asymmetry between the SQUID arms with the respective gate-voltages we show that we can achieve a SDE up to 30\%. This comes close to the maximum theoretically predicted value~\cite{Souto2022}.

In Sec.~\ref{sec:Device} we present the device geometry, the experimental set up and the basic characterization of the individual JJs.
The non-reciprocal character of the dc-SQUID with JJs having a non-sinusoidal CPR is then shown in Sec.~\ref{sec:Josephson Diode Effect}. We also define an analytical framework with which we are able to distinguish possible origins of the JD effect.
Finally, we discuss the measured gate tunability of the diode efficiency in Sec.~\ref{sec:Gate Tunable Diode Efficiency} and end with the conclusion in Sec.~\ref{sec:Conclusion}.

\section{Device and Basic Properties}
\label{sec:Device}
\vspace{-0.3cm}

\begin{figure}[ht]
	\centering
	\includegraphics[width=\columnwidth]{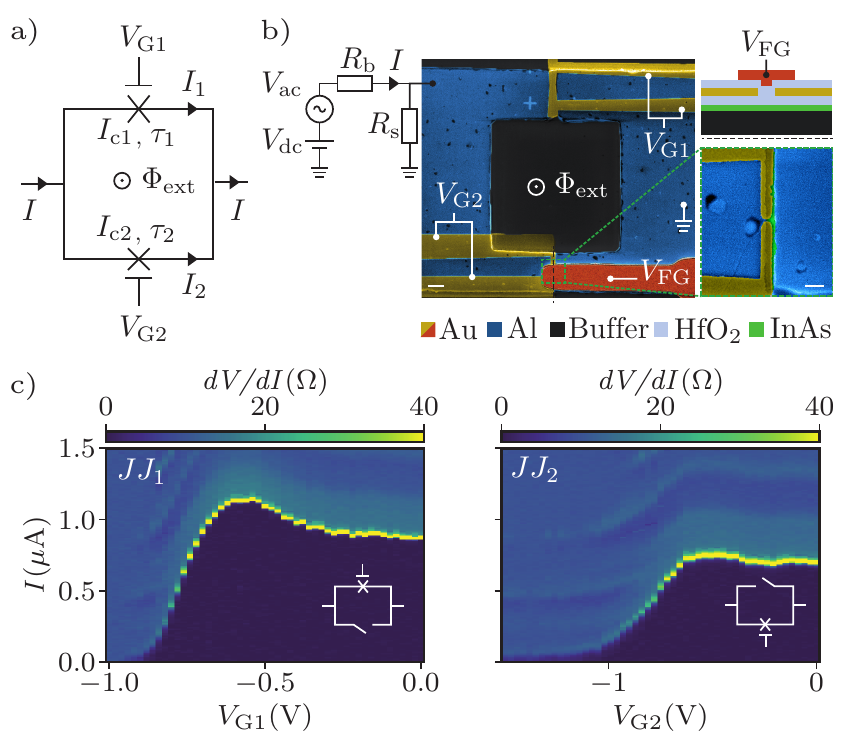}
	\caption{(a) Circuit schematic of a dc SQUID threaded by the external flux $\Phi_{\mathrm{ext}}$, formed by two gate tunable JJs with non-sinusoidal CPRs
    with critical currents $I_{\textrm{c1}}$, $I_{\textrm{c2}}$, and transparencies $\tau_1$, $\tau_2$. (b) False-color electron micrograph of the device. The loop consists of a $10$~nm Al film (blue) grown on top of an InAs 2DEG (green). The JJs are defined by selectively removing the Al over 150~nm long stripes on each branch of the loop. Electrostatic gates (yellow and orange) tune the charge carrier density in the junction. We use 15~nm of HfO$_2$ (light blue) as a gate dielectric. On the right, a zoom-in of JJ$_2$ is shown before adding the $\mathrm{FG}$. On top, we show a cross-sectional schematic of the gate configuration of JJ$_2$ along the dashed black line. The scale bar in the main figure is 1~$\mu$m and in the zoom-in it is 300~nm. Dc and ac current bias are defined through the voltage drop over a large series resistor with value $R_b=1$~M$\Omega$. The SQUID is shunted to ground with  a parallel resistor of value $R_s=10$~$\Omega$. (c) Differential resistance of JJ$_1$ (left) and JJ$_2$ (right) as a function of gate voltage and current bias. While one junction is being measured, the other is pinched-off. The top junction has a slightly higher critical current due to the different channel widths of W$_1$=3~$\mu$m and W$_2$=2.5~$\mu$m.}
	\label{fig:Fig1}
\end{figure}

The circuit diagram of the device is shown in Fig.~\ref{fig:Fig1}(a) and a coloured electron-microscopy picture is presented in Fig.~\ref{fig:Fig1}(b).
%
The circuit consists of a dc SQUID formed by two planar JJs realized in a shallow InAs 2DEG proximitized by Al layer. The 2DEG is obtained from a quantum well grown on an InP substrate embedded in In$_{0.75}$Ga$_{0.25}$As layers of which the top layer is $10$~nm thick. The stack is terminated with an in-situ grown $10$~nm thin Al layer inducing superconductivity in the 2DEG. The SQUID loop and the leads are defined by etching the Al and, additionally, $300$~nm deep into the semiconductor stack. The top and bottom Josephson junctions (JJ$_1$ and JJ$_2$) in the two branches of the loop are formed by selectively removing the Al in the form of stripes with length $L=150$~nm and width $W_1=3$~$\mu$m and $W_2=2.5$~$\mu$m.\\
A set of gates, $\mathrm{G1}$, $\mathrm{G2}$ and $\mathrm{FG}$, are used to tune the critical current of the junctions by applying appropriate gate voltages $V_{\mathrm{G1}}$, $V_{\mathrm{G2}}$ and $V_{\mathrm{FG}}$. They are made of two Ti/Au layers, isolated from the Al and from each other by hafnium dioxide (HfO$_2$) layers. $V_{\mathrm{G1}}$ extends over the whole width of JJ$_1$, while $V_{\mathrm{G2}}$ is shaped to gradually deplete JJ$_2$ laterally, creating a Superconducting Quantum Point Contact (SQPC). An additional gate, $V_{\mathrm{FG}}$, can be use to fine tune the charge carrier density in the SQPC. However, throughout the experiment the QPC functionality is not used and $V_{\mathrm{FG}}$ is kept at $0$~V.\\

Our setup sources a current using a $1$~M$\Omega$ resistor in series to a dc voltage superposed by a small ac component with frequency $f$ = 17.7~Hz, supplied by a lock-in amplifier. The ac component has an amplitude of $5$~nA.
The SQUID is additionally shunted at the source to ground with a resistor $R_S=10$~$\Omega$ directly placed on the sample holder.
This shunt resistor has two purposes: a) it limits the maximum voltage that appears over the junction in the normal state, and thus, the heating; and (b) it adds damping to the device avoiding hysteretic switching when assessing the critical current in experiments.
We measure the differential resistance of the shunted device using a voltage amplifier and lock-in techniques. In all plots where a measured differential resistance $dV/dI$ is shown the shunt resistor was not subtracted. The measurements presented in the following were obtained with the SQUID device operating in a dilution refrigerator with a base temperature of $\sim 50$~mK.\\

In Fig.~\ref{fig:Fig1}(c) we show the measured differential resistance of JJ$ _1 $ (left) and JJ$ _2 $ (right) as a function of gate voltage and bias current. In the following, we approximate the critical current $I_{\textrm{ci}}$ of the $\textrm{i}$th-junction, $\textrm{i}=\{1,2\}$, by the current bias value at which the maximum value in differential resistance is measured. Here, the bias current is swept from zero to $1.5$~$\mu$A, looking at transitions from the super\-conducting to the normal state. From the measurements we extract $I_{\textrm{ci}}(V_{\mathrm{Gi}})$. The critical current of both junctions can be tuned from a few nA close to pinch-off at negative gate-voltages $V_{\mathrm{G(1,2)}}\lesssim -1$~V to approximately 1~$\mu$A.
The key features of these hybrid semiconducting-super\-conducting JJs are the gate tunable critical current and the non-sinusoidal CPR.

In the short-junction limit, i.e. for junctions with a length $L$ shorter than the super\-conducting coherence length $\xi$ in the normal metal, the zero temperature limit of the super\-current $I(\varphi)$ is given by~\cite{Koops1996}:
\begin{equation}
  I(\varphi)=\sum_j\left(\frac{\tau_j e\Delta}{\hbar}\right)\frac{\sin(\varphi)}{\sqrt{1-\tau_j \sin^2(\varphi/2)}}.
\end{equation}
Here, $\tau_j$ is the transmission probability per channel $j$. In multichannel devices with disorder, a universal distribution function of transmission eigenvalues was obtained~\cite{Kulik1975,Dorokhov1984,Nazarov1994,Beenakker1997}. The distribution is bimodal with many low transmissive channels that contribute little to the current, but also with some channels having a transmission probability close to 1. These high-transmissive channels lead to the overall non-sinusoidal character. This is approximated with an effective (but constant) transmission probability $\tau^*$ per channel and written as a single channel non-sinusoidal CPR given by:
\begin{equation}
  I(\varphi)=\frac{I_{\textrm{c}}}{A_N}\frac{\sin(\varphi)}{\sqrt{1-\tau^*\sin^2(\varphi/2)}}.
\label{eq:CPR}
\end{equation}
For the later discussion of the measurements the critical current $I_{\textrm{c}}$ of the junction and a unit-less normalization parameter $A_N$ are introduced. The ratio $I_{\textrm{c}}/A_N$ is given by $N\tau^*e\Delta/\hbar$ with $N$ the number of channels. Note, for the single junction we have $I(-\varphi)=-I(\varphi)$ and thus $I_{\textrm{c}}^{+}=I_{\textrm{c}}^{-}=I_{\textrm{c}}$. It is also seen that for small values of $\tau^*$ the CPR approaches a sinusoidal dependence.
From experimental $I(\varphi)$ curves, we deduce the critical current $I_{\textrm{c}}$ of each junction, $\tau^*$ and $A_N$. Note, that only two parameters are independent.

As shown in Fig.~\ref{fig:Fig1}(a) the total super\-current $I$ across the SQUID is the sum of the currents flowing in both branches $I_1$ and $I_2$ through the two JJs:
\begin{equation}
  I(\varphi_1,\varphi_2)=I_1(\varphi_1)+I_2(\varphi_2).
\label{eq:ISquid1}
\end{equation}
The two junctions are described by $I_{\textrm{c1}},I_{\textrm{c2}}$ and $\tau^*_1, \tau^*_2$. The uniqueness of phase around the loop leads to the so-called fluxoid relation (modulo $2\pi$)
\begin{equation}
  \varphi_1 - \varphi_2 = 2\pi \Phi_{\mathrm{ext}}/\Phi_0=\varphi_{\mathrm{ext}},
\label{eq:fluxoid}
\end{equation}
where $\Phi_{\mathrm{ext}}$ denotes the externally induced flux, $\Phi_0=~h/2e$ the super\-conducting flux quantum and $\varphi_{\mathrm{ext}}$ the respective phase. In this form of the fluxoid relation the loop inductance has been neglected. For a finite loop inductance there is an additional flux contribution which depends on the currents $I_1$ and $I_2$ flowing in each arm. It has been shown that asymmetric loop inductances can also induce a super\-conducting SDE~\cite{Barone1982,Tesche1977,Clarke2004}. To estimate the role of loop inductances in our experiment we perform a full analysis with equations given in the appendix, specifically in App.~\ref{sec:Model Including Loop Inductances}. Taking Eq.~\ref{eq:ISquid1} and Eq.~\ref{eq:fluxoid} together yields an effective super\-conducting junction with a CPR
\begin{equation}
  I(\varphi)=I_1(\varphi)+I_2(\varphi-\varphi_{\mathrm{ext}}) \text{.}
\label{eq:ISquid2}
\end{equation}
For a simple sinusoidal CPR, the addition of the two terms yields a $\varphi_0$-junction without a SDE, even when the two JJ have different critical currents. In contrast, in the presence of higher order harmonics, which appear for a non-sinusoidal CPR, constructive and destructive interference effects, acting opposite for the two current bias directions, give rise to unequal critical currents  $ I_{\textrm{c}}^+\neq I_{\textrm{c}}^-$, and thus to a SDE~\cite{Souto2022,Fominov2022}.\\

\section{Josephson Diode Effect}
\label{sec:Josephson Diode Effect}
\vspace{-0.3cm}

\begin{figure}[ht]
	\centering
	\includegraphics[width=\columnwidth]{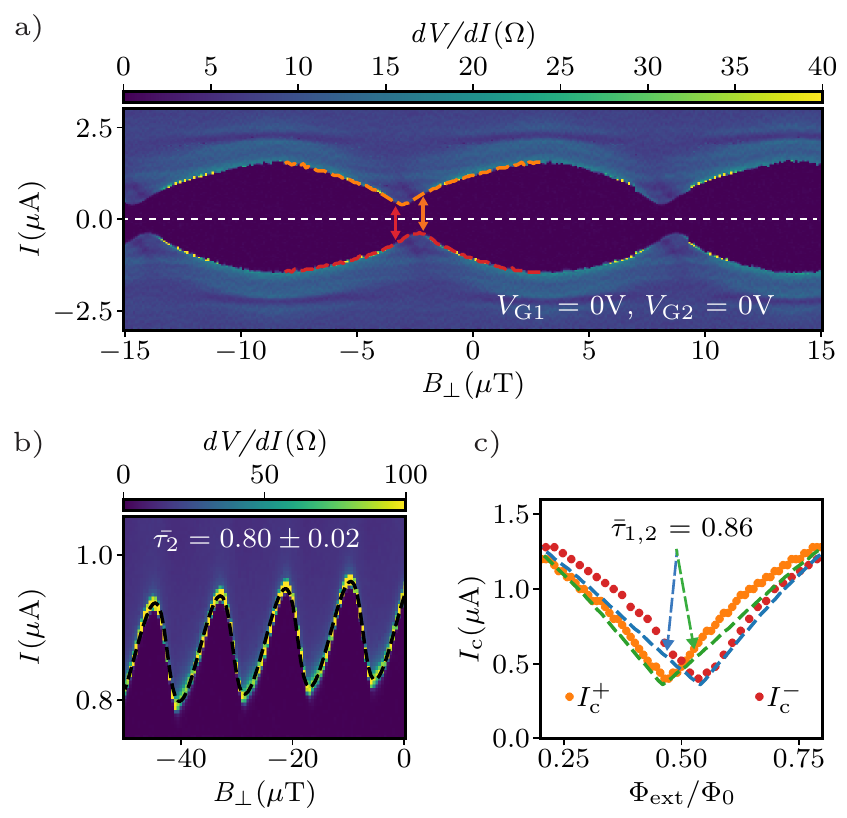}
	\caption{(a) SQUID oscillations with $V_{\mathrm{G1}} = V_{\mathrm{G2}} = 0$. The critical current $I_{\textrm{c}}^+$
      and the retrapping current $I_{\textrm{r}}^-$ over one flux period are highlighted in orange and red respectively. At fixed magnetic field, the absolute value of the critical current in the two sweep directions is not the same. This is best seen in the region $-5 < B_{\perp} < 0$~$\mu$T  with a visible example taken at the red and orange arrows, where the SDE has a magnitude of $\sim 23$~\%.
      (b) Measurement for a strongly asymmetric SQUID setting with $V_{\mathrm{G1}}=0$~V and $V_{\mathrm{G2}}=-1.1$~V. Here, the junction with the large critical current JJ$_1$ serves as the reference junction. As a consequence, the critical current as a function of flux now reflects the CPR of the weaker junction JJ$_2$. The CPR is strongly non-sinusoidal and a fit (black dashed line) yields $\tau^*_2=0.8$.
      (c) Plot of the extracted $I_{\textrm{c}}^+$ (orange) and $I_{\textrm{c}}^-$ (red)
      taken from the measurement in (a) and from a measurement where we sweep the current bias from positive to negative values (see SM). The dashed two curves (green and blue) show simplified model fits with $\tau^*_{1,2}=0.86$ and the critical currents of the junctions taken from Fig.~\ref{fig:Fig1}(c)
     }
	\label{fig:Fig2}
\end{figure}

Figure~\ref{fig:Fig2}(a) shows the differential resistance of the SQUID as a function of current bias and perpendicular magnetic field $B_{\perp}$, the latter providing the flux $\Phi_{\mathrm{ext}}$ through the SQUID loop. We have chosen a gate configuration with $V_{\mathrm{G1}}=V_{\mathrm{G2}} = 0$~V for which the two critical currents are similar: $I_{\textrm{c1}}=0.87$~$\mu$A and $I_{\textrm{c2}}=0.67$~$\mu$A. A clear SDE is visible. For example, at the place of the orange arrow, we obtain $I_{\textrm{c}}^+=0.64$~$\mu$A and $I_{\textrm{c}}^-=0.4$~$\mu$A.

In this experiment, the current bias is swept from negative to positive values. This means that we measure the positive switching current $I_c^+$, but on the negative side, we actually measure what is called the retrapping current $I_r^-$ where the device switches from the normal to the super\-conducting state. Due to dissipation, the junction can overheat in the normal state giving rise to a hysteresis between the switching and retrapping currents with the retrapping current being smaller in magnitude than the switching current.  This would result in an artificial SDE. To exclude this, we have measured the same plot as in Fig.~\ref{fig:Fig2}(a) but sweeping now from positive to negative bias currents. The comparison shows, see App.~\ref{sec:Retrapping versus Switching Current}, that the hysteresis between retrapping and switching currents is small and can be neglected. Physically, this is the case thanks to the low shunt resistant of $R_s=10$~$\Omega$ which limits the voltage over the junction to $< 25$~$\mu$V, and thus, limits the heating.

Another strong argument against an artificial effect is seen in Fig.~\ref{fig:Fig2}(a) when one looks at the switching values at the place of the red arrow, where $I_{\textrm{c}}^+=0.44$~$\mu$A and $I_{\textrm{r}}^-=0.6$~$\mu$A. Here, the sign of the SDE is reversed, $I_{\textrm{c}}^+ < I_{\textrm{r}}^-$. This cannot be explained by a hysteresis between the switching and retrapping currents, since the retrapping current should always be smaller than the switching current.

As introduced before, a contribution from loop inductances may generate the SDE, too, if the loop inductances in the two arms are different. Applying finite element simulations, App.~\ref{sec:Loop Inductance}, we obtain $L_1 \approx 39$~pH and $L_2\approx 44$~pH.  The relative phase shift between the two SQUID arms due to the loop inductances at a bias current $I = \SI{1}{\micro\ampere}$ is only $\frac{2\pi}{\Phi_0}(L_2-L_1)I \sim  \SI{0.03}{\radian}$, and gives a small contribution to the SDE. We properly simulate the effect of the loop inductances on the critical current of the SQUID in App.~\ref{sec:Comparison to Diode Effect by Loop Inductance} and find that the loop inductances alone cannot explain the observed SDE in our experiment.

We also note that the measured CPR of the SQUID in Fig.~\ref{fig:Fig2}(a) is periodic with a periodicity of $11.6$~$\mu$T. Since this should correspond to an added flux quantum $\Phi_0$ in the area $A_{h}$ of the inner SQUID hole, we obtain for $A_h=175$~$\mu\textrm{m}^2$. This is approximately a factor of $2.3$ bigger than the geometrical area defined by the etched square-shaped hole of size $75$~$\mu\textrm{m}^2$. This discrepancy can be attributed to the flux-focussing effect~\cite{Granata2016}. The magnetic field above the super\-conductor is screened by the Meissner effect leading to an enhanced magnetic field within the inner hole. The enhancement factor can be estimated by the ratio of the outer superconducting loop area of $\approx 150$\,$\mu\text{\rm m}^2$ relative to $A_{h}$, which yields a factor of $2$ in good agreement with the experiment.


%

In a sufficiently asymmetric SQUID configuration one can measure the CPR of the weak junction alone~\cite{Rocca2007}.
Figure~\ref{fig:Fig2}(b) shows a measurement of the CPR of a single junction, obtained during the same cool-down. Here, $V_{\mathrm{G1}}=0$~V and $V_{\mathrm{G2}}=-1.1$~V so that the current in JJ$_1$ is large $\sim 0.9$~$\mu$A and in JJ$_2$ it is small $\sim 0.1$~$\mu$A. In such a situation JJ$_1$ acts as reference junction and the critical current of the weak junction JJ$_2$ can be obtained from Eq.~\ref{eq:ISquid2} as
\begin{eqnarray}
  I_{\textrm{c}}^+=\max_\varphi(I_1(\varphi)+I_2(\varphi - \varphi_{\mathrm{ext}})) \\
  I_{\textrm{c}}^+(\varphi_{\mathrm{ext}})\simeq I_{\textrm{c1}}+I_2(\tilde{\varphi}_1 - \varphi_{\mathrm{ext}}),
\end{eqnarray}
where $\tilde{\varphi}_1$ is the phase value for which JJ$_1$ has its maximal value $I_{\textrm{c1}}$. Hence, we see that under the condition that the reference junction dominates, we obtain the phase dependence of the critical current of the weak junction from the flux dependence of the critical current of the SQUID. Applying Eq.~\ref{eq:CPR} to fit the measured data yields for the effective transmission probability $\tau^*=0.8 \pm 0.02$. This is a large value, showing that the CPR is strongly non-sinusoidal, something that is visibly seen in the graph of Fig.~\ref{fig:Fig2}(b). If one makes use of the universal bimodal distribution function of transmission eigenvalues to determine $\tau^*$~\cite{Kulik1975,Dorokhov1984,Nazarov1994,Beenakker1997}, one obtains $\tau^*=0.866$. Including different devices nominally fabricated the same way, we always find a large effective transmission value of order $\sim 0.8$ in agreement with theoretical expectations for a multi\-channel disordered junction in the short junction limit.

In Fig.~\ref{fig:Fig2}(c) we compare the oscillations of $I_{\rm c}^{+}$ and $I_{\rm c}^{-}$ as a function $\Phi_{\rm ext}$
with the simplified model of Eq.~\ref{eq:ISquid2}.
%
We take the measured critical currents of the two junctions as input parameters, i.e. $I_{\textrm{c1}}=0.87$~$\mu$A and $I_{\textrm{c2}}=0.67$~$\mu$A, and assume $\tau^*_1=\tau^*_2=\tau^*$ as a single fitting parameter. The best agreement is obtained for $\tau^*=0.86$. We note, that a similar model calculation based only on loop inductances barely matches the measurement. It is shown as a comparison in App.~\ref{sec:Comparison to Diode Effect by Loop Inductance}. \\

The fits for $I_{\textrm{c}}^+$ (green) and $I_{\textrm{c}}^-$ (blue) reproduce the relative shift along the flux axis very well. The shape of the curves is, however, not reproduced so well. In the region $\Phi_{\mathrm{ext}}/\Phi_0 \in [0.25,0.5]$ and $\Phi_{\mathrm{ext}}/\Phi_0 \in [0.5,0.75]$ respectively, the measured $ I_{\textrm{c}}^+$ and $ I_{\textrm{c}}^-$ curves are higher than what is obtained with the model. Deviations between the experimental and the modelled curves could be attributed to the choice of CPR used in the model. First, we considered an average transparency instead of a distribution of transparencies. Second, the expression of the current carried by the Andreev bound states could be different from Eq.~\ref{eq:CPR}, since our junctions could be in a regime intermediate to the short and long junction limit. And, in the third place, spin-orbit effects may affect the CPR, too. For junctions of similar length in the same material system, it has been shown that spin-orbit interaction splits the ABS into spinful states with different dispersion relations~\cite{Tosi2019}.
Noticeably, the experiment indicates that these deviations result in an increase of the SDE compared to what is predicted by the simple model.

Having established that a SDE appears in a SQUID with junctions having a non-sinusoidal CPR with asymmetry, we summarize in Table~\ref{table-conditions-for-DE} the necessary conditions for the SDE (DE). To describe the asymmetry we introduce two asymmetry parameters $\alpha$ and $\beta$ for the critical currents and the effective transmission probabilities, respectively:
\begin{equation}\label{eq:asymmetry}
  \alpha =\frac{I_{\textrm{c1}}-I_{\textrm{c2}}}{I_{\textrm{c1}}+I_{\textrm{c2}}} \textrm{\hspace{5mm} and \hspace{5mm}}
  \beta =\frac{\tau^*_{1}-\tau^*_{2}}{\tau^*_{1}+\tau^*_{2}}
\end{equation}
\\

\begin{table}
  \label{table-conditions-for-DE}
  \caption{Conditions for obtaining a SDE (DE). An extended table that includes the loop inductances can found in App.~\ref{sec:Conditions for a Diode Effect in a SQUID Device}. The first column is used to distinguish the classical sinusoidal CPR ($\tau^*=0$) from a strongly skewed CPR described by a highly transmissive ballistic JJ with an effective transmission probability $\tau^*>0$. $\alpha$ ($\beta$) denotes the asymmetry in critical currents (transmission probabilities) of the two junctions.\\}
  \begin{minipage}{0.4\textwidth}
  \begin{ruledtabular}
    \begin{tabular}{|c|c|c|c|}
      $\tau^*$ & $\alpha$ & $\beta$ & SDE \\
      \hline
      $0$ & $0$ & n.a. & no  \\
      \hline
      $0$ & $\neq 0$ & n.a. & no \\
      \hline
      $\neq 0$ & $0$ & $0$ & no  \\
      \hline
      $\neq 0$ & $0$ & $\neq 0$ & yes \\
      \hline
      $\neq 0$ & $\neq 0$ & $0$ & yes \\
      \hline
      $\neq 0$ & $\neq 0$ & $\neq 0$ & yes
    \end{tabular}
  \end{ruledtabular}
  \end{minipage}
\end{table}
\vspace{-0.3cm}

An extended table, which also considers the effect of loop inductances, is presented in App.~\ref{sec:Conditions for a Diode Effect in a SQUID Device}. It shows that the diode effect appears when the SQUID arms are asymmetric. The only exception is for sinusoidal JJs, where an asymmetry in the critical currents in not enough to produce a diode effect.


\section{Gate Tunable Diode Efficiency}
\label{sec:Gate Tunable Diode Efficiency}
\vspace{-0.3cm}

The SDE can be quantified via the diode efficiency, defined as
\begin{equation}
\eta = \frac{I_{\textrm{c}}^+ - I_{\textrm{c}}^-}{I_{\textrm{c}}^+ + I_{\textrm{c}}^-} \mathrm{~.}
\end{equation}

\begin{figure}[ht]
	\centering
	\includegraphics[width=\columnwidth]{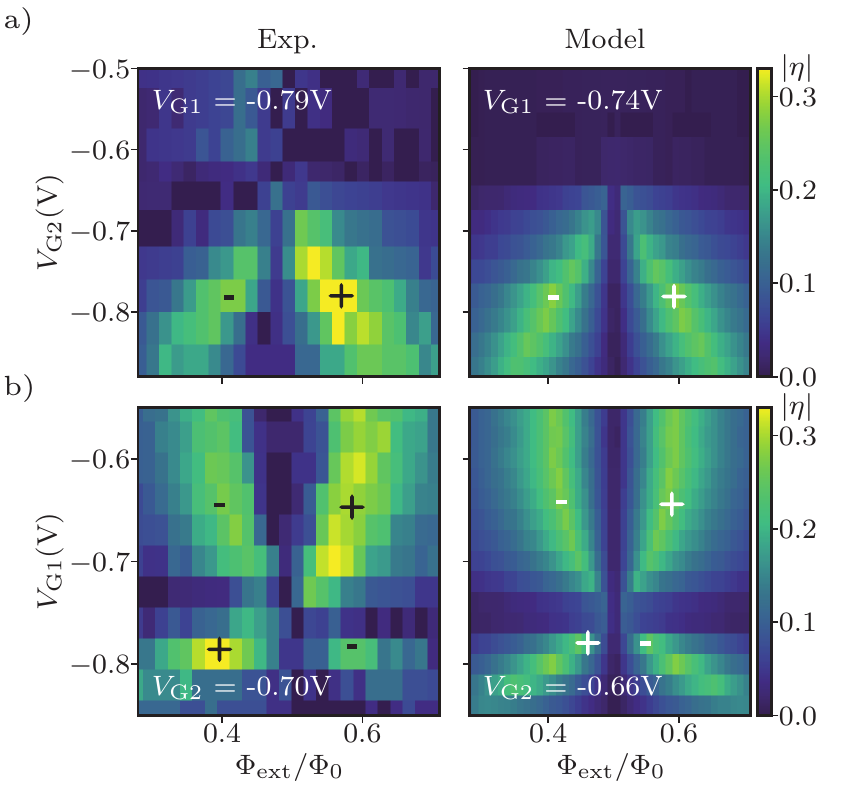}
	\caption{Magnitude of the diode efficiency $|\eta|$ as a function of external flux $\Phi_{\mathrm{ext}}$ for different gate configurations as obtained from
     the measurements (left) and as calculated from the model (right). The sign of $\eta$ is indicated on the visible lobes with $+$ and $-$.
     The model takes into account the numerically simulated loop inductances, their asymmetry, and the values $I_{c(1,2)}$ of the two junctions obtained from the measurements in Fig.~\ref{fig:Fig1}(c). The JJ transparencies were fixed to $\tau^*_1 = \tau^*_2 = 0.86$.
     (a) $|\eta|$ as a function of $V_{\mathrm{G2}}$ at fixed $V_{\mathrm{G1}}$, and (b) $|\eta|$ as a function of $V_{\mathrm{G1}}$ at fixed $V_{\mathrm{G2}}$. Note, that for $\Phi_{\textrm{ext}}/\Phi_0=0.5$, which equals $\varphi_{\mathrm{ext}}=\pi$, $\eta=0$ independent on any other parameters.
     }
	\label{fig:Fig3}
\end{figure}

In Fig.~\ref{fig:Fig3}, we show the magnitude of the diode efficiency $|\eta|$ as a function of external flux $\Phi_{\mathrm{ext}}/\Phi_0$ for different gate configurations as obtained from the experiment (left) and as calculated from the model (right). In the model, we make use of the relation between critical current and gate voltage of the individual junctions $I_{\textrm{ci}}(V_{\mathrm{Gi}})$ and use these values as input parameters in the first approximation. We also use the simulated loop inductance values from which we obtain the phase response due to screening $\varphi_{\mathrm{L}}=4\pi\bar{I_{\textrm{c}}}\bar{L}/\Phi_0$, the loop inductance asymmetry $\gamma=(L_1-L_2)/(L_1+L_2)$ with $L_1$, $L_2$, and $\bar{I}_{\textrm{c}}$ and $\bar{\tau}^*$ the respective mean values. We assume that the effect of the gate voltage is mainly to change the critical current value $I_{\textrm{ci}}$ through the number of channels $N$, while $\tau^*_{\textrm{i}}$ roughly stays constant. We fix $\tau^*_1=\tau^*_2=0.86$, but we note that the calculated $\eta$ plot is insensitive if one varies $\tau^*_2$ between $0.8$ and $0.9$. \\

In Fig.~\ref{fig:Fig3}(a), we plot $|\eta|$ for different values of $V_{\mathrm{G2}}$ at fixed $V_{\mathrm{G1}}$. Both in the experiment and in the model, $|\eta|$ drops for $-0.7 < V_{\mathrm{G2}} <-0.5$~V. As seen in Fig.~\ref{fig:Fig1}(c), this corresponds to a gate configuration with $I_{\textrm{c1}}\approx I_{\textrm{c2}}$, so that $\alpha \approx 0$. As expected, the absence of critical current asymmetry decreases the diode efficiency.
To obtain in the model the same diode efficiencies $\eta$ as measured, we had to increase the critical current of JJ$_1 $. In the experiment, we had $ V_{\mathrm{G1}}$ fixed at $-0.79$~V, which would correspond to $I_{\textrm{c1}}=470$~nA. However, in order to match the model with the data, we had to use $710$~nA, corresponding to $V_{\mathrm{G1}}=-0.74$~V, as indicated in the top left corner of the figure. Without this correction, the measured $|\eta|$ values would have been larger than what the model predicts. We attribute this difference in gate voltage to gate-jumps that occur from time-to-time. We note, that there are days between the measurements in Fig.~\ref{fig:Fig1}(c) and in Fig.~\ref{fig:Fig2}(a)

In Fig.~\ref{fig:Fig3}(b) we show the dependence of $|\eta|$ as a function of $V_{\rm G1}$ at fixed $V_{\rm G2}=\SI{-0.7}{\volt}$. As before, to match the model to the experiment, we had to increase $ I_{\textrm{c2}}$ from the initially measured value of $590$~nA at $V_{\mathrm{G2}}$ to $650$~nA, which correspond to $I_{\textrm{c2}}$ measured at $V_{\mathrm{G2}}=-0.66$~V.

Both in the experiment and in the model one can observe the typical butterfly pattern of $\eta $ as predicted in Ref.~\cite{Souto2022}. The two arms of maximum $|\eta|$ meet at the point of minimum asymmetry at $\Phi_{\mathrm{ext}}/\Phi_0 =0.5$ for $V_{\mathrm{G2}}\approx -0.65$~V and $V_{\mathrm{G1}}\approx -0.75$~V for (a) and (b) respectively, where $\eta$ drops to $0$.

The model qualitatively reproduce the gate dependence of the diode efficiency very well. We obtain a maximum $|\eta|$ of $\simeq 0.3$ from the experiment. This $30\%$ efficiency is much larger than what has previously been obtained in a SQUID with asymmetric loop inductance~\cite{Paolucci2023}. Taking a SQUID model with a single channel JJ junction, we numerically find for the maximum efficiency $\eta=0.37$. This is obtained for $\tau_1=1$ and $\tau_2=0.75$ or the reversed. This could be achieved by combining a single channel ballistic $\tau=1$ Josephson junction realized in atomic contacts~\cite{Rocca2007} with a semiconductor-super\-conductor hybrid device as we have discussed here.

\section{Conclusion}
\label{sec:Conclusion}
\vspace{-0.3cm}
In conclusion, we have investigated the origin of the Super\-conducting Diode Effect (SDE) in a super\-current interferometer realized in a proximitized InAs quantum well stack. We show that in such a system the SDE can originate from the non-sinusoidal character of the JJs, and hence, reflecting a subtle interference between higher-order harmonics of the CPRs of the individual JJs. In addition to higher harmonics, an asymmetry either in the composition of the Fourier components in the CPR or in the critical current of the two JJ, and a finite flux bias $\varphi_{\mathrm{ext}}\not = \{0,\pi \}$ is required to obtain a SDE. These later conditions ensure that time-reversal symmetry and inversion symmetry are both broken.
A similar conclusion was drawn by a recent experimental study in three terminal devices, where a SDE was realized~\cite{Gupta2023}. Further, during the reviewing process we got aware of a similar study in a dc SQUID realized in a Ge quantum well structure~\cite{Valentini2023}.
Future directions include the possibility to concatenate more SQUIDs in parallel in order to further increase the diode efficiency as was proposed in Ref.~\cite{Souto2022}.

\begin{acknowledgments}
\vspace{-0.3cm}
  We thank  C. M. Marcus for his support in initiating this work and collaboration. This research was supported by the Swiss National Science Foundation through grants No 172638 and 192027, and the QuantEra project SuperTop. We further acknowledge funding from the European Union’s Horizon 2020 research and innovation programme, specifically a) from the European Research Council (ERC) grant agreement No 787414, ERC-Adv TopSupra, (b) grant agreement No 828948, FET-open project AndQC, c) grant agreement 847471, project COFUND-QUSTEC, and d) grant agreement 862046, project TOPSQUAD. Constantin Schrade acknowledges support from the Microsoft Corporation and Christian Sch{\"o}nenberger from the Swiss Nanoscience Institute (SNI). All data in this publication is available in numerical form at: \url{https://doi.org/10.5281/zenodo.7733057}.
\end{acknowledgments}

\appendix
\section{Fabrication \& Measurement Set-up}
\label{sec:Device Setup}
\vspace{-0.3cm}

The wafer used in this experiment was grown by molecular beam epitaxy (MBE). The stack consists from bottom to top of an InP substrate, a 1-$ \mu $m-thick buffer realized with In$_{1-x}$Al$_{x}$As alloys, a 4~nm In$_{0.75}$Ga$_{0.25}$As bottom barrier, a 7~nm InAs layer, a 10~nm In$_{0.75}$Ga$_{0.25}$As top barrier, two monolayers of GaAs acting as stop etch layer, and 10~nm of Al deposited $ in~situ $ without breaking the MBE vacuum. The two-dimensional electron gas is characterized from a Hall bar devices and shows a peak electron mobility of $\mu=12'000~\textrm{cm}^2\textrm{V}^{-1}\textrm{s}^{-1}$ for an electron density of $ 16~\textrm{x}~10^{11}~\textrm{cm}^{-2}$, corresponding to an electron mean free path of $ l_{\textrm{e}} \approx 230$~nm.\\

\begin{figure}[h]
	\centering
	\includegraphics[width=\columnwidth]{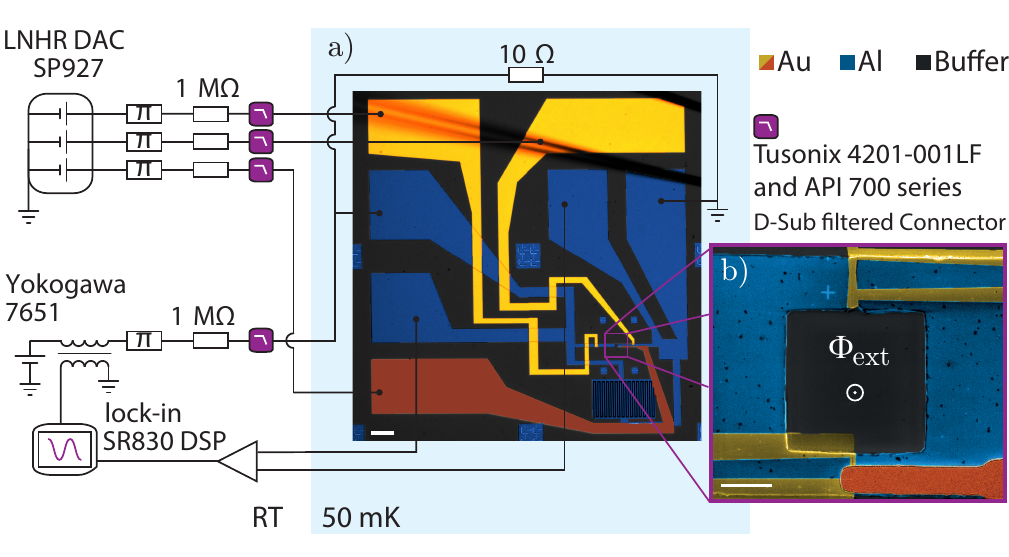}
	\caption{(a) False color optical image of the full device together with a sketch of the measurement setup. The scale bar is $100$~$\mu$m.
       (b) Zoom-in over the SQUID showing the loop area threaded by the external flux $\Phi_{\mathrm{ext}}$. The electron density in the junction region is tuned via a set of gates coloured in yellow and brown. The scale bar is $3$~$\mu$m}
	\label{fig:S1}
\end{figure}

The device is fabricated using standard electron beam lithography techniques. The MESA is electrically isolated by first removing the top Al film with Al etchant Transene D, followed by a deep III–V chemical wet etch with H$ _2 $O:C$ _6 $H$ _8 $O$ _7 $:H$ _3 $PO$ _4 $:H$ _2 $O$ _2 $ (220:55:3:3). Next, the Al film on the mesa is selectively etched with Al etchant Transene D to define the planar JJ. Electrostatic gates are made of two Ti/Au layers, isolated from the Al and from each other by hafnium oxide (HfO$_2$) layers grown by atomic layer deposition (ALD) at a temperature of $90$~°C over the entire sample. The first layer of gates is made of electron-beam evaporated Ti/Au ($5$~nm/$25$~nm) on top of $15$~nm HfO$_2$. Connections to the external circuit are obtained by evaporating Ti/Au ($5/85$~nm) leads at $ \pm $17° to overcome the MESA step. A second layer of gates, made of angle-evaporated Ti/Au ($5/85$~nm), is patterned on top of $25$~nm of HfO$_2$.\\

Measurements are carried out in a Triton 200 cryogen-free dilution refrigerator with a base temperature of $\approx 50$~mK. An overview of the measurement set-up is shown in Fig.~\ref{fig:S1}. The setup sources a current using a $1$~M$\Omega$ resistor in series to a dc voltage source on which a small ac component with frequency $f$ = 17.7~Hz, supplied by a lock-in amplifier, is superposed. This current is applied to the source contact of the SQUID on the left with the drain contact on the right side galvanically connected to ground. The SQUID is shunted at the source to ground with a resistor $R_S=10$~$\Omega$. This shunt resistor is directly placed on the sample holder. In addition, a finger capacitance of $ \approx 0.7~$pF is patterned in parallel to the SQUID (lower right of the optical image). The original purpose of the capacitance was to increase the quality factor of the Josephson junctions. However, its effect is negligible, since the capacitance provided by the leads is larger. We measure the differential resistance of the shunted device using a voltage amplifier and lock-in techniques. The flux through the SQUID is generated by a vector magnet.\\

\section{Estimation of Loop Inductances}
\label{sec:Loop Inductance}
\vspace{-0.3cm}

\begin{figure}[h]
	\centering
	\includegraphics[width=\columnwidth]{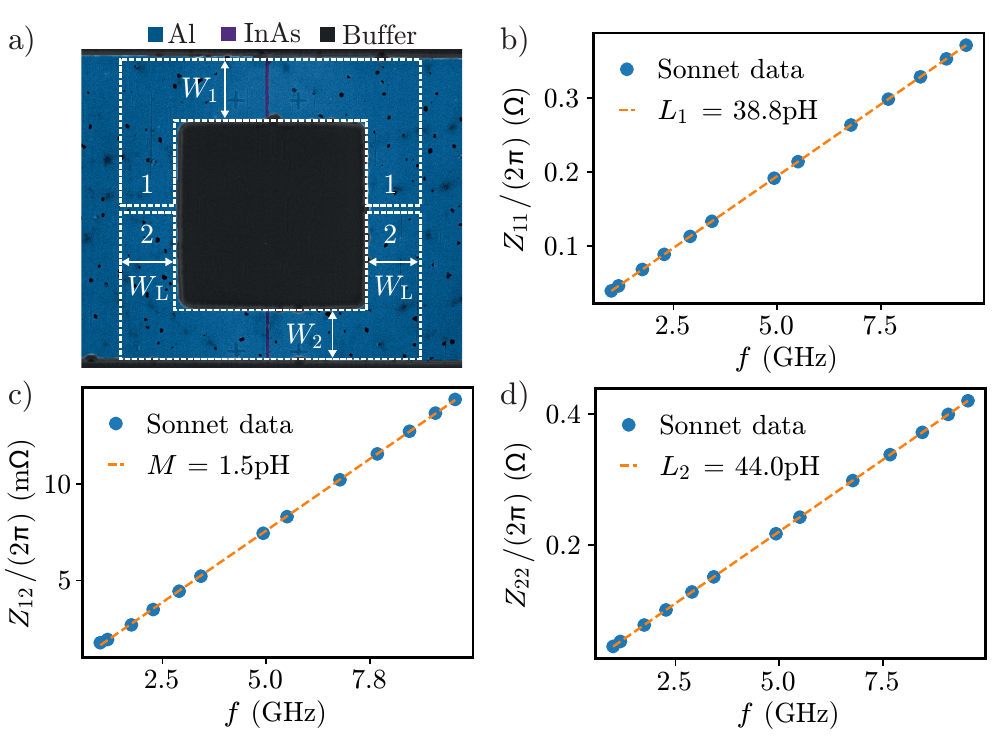}
	\caption{Sonnet simulations of the loop inductances. The super\-conducting loop is segmented into an upper (lower) branch $1$ ($2$)
      indicated by the white dashed boxes. The respective width are $W_1=3$~$\mu$m, $W_2=2.5$~$\mu$m and $W_L=2.75$~$\mu$m. The two inductances $L_1$, $L_2$ and the mutual inductance $M$ are deduced from the slope of the frequency dependent two-port impedances. It is seen that $M \ll L_{1,2}$ and that there is a small asymmetry of $\sim 6$~$\%$ in the loop inductances.}
	\label{fig:S2}
\end{figure}

In the following we will detail the evaluation of the inductance of the loop branches. The loop geometry is defined as indicated by the white dashed lines in Fig.~\ref{fig:S2}(a). The width of the two branches corresponds to the junctions width in the upper and lower path, $W_{\textrm{1}}$ = $3$~$\mu$m, $W_{\textrm{2}}$ = $2.5$~$\mu$m, and the width on the left and right sides it is set equal to $W_{\textrm{L}} = (W_{\textrm{1}}+W_{\textrm{2}})/2$ = $2.75$~$\mu$m. In reality there is no lateral confinement in the superconductor. Hence, the artificial confinement increases the inductance values so that the simulated inductances for this geometry yield upper bounds to the inductances of the device. With finite-element simulations performed in Sonnet, we compute the two-port impedances $Z_{\textrm{i,k}}$ with $\textrm{i,k} \in \{1,2\}$ for different frequencies. The impedance is evaluated between two sets of floating co-calibrated ports, positioned on the left and right side of the loop. In the simulation we use InP as a substrate, with a relative dielectric constant $\epsilon_r  = 12.4$. The kinetic inductance of the Al film is evaluated by measuring the temperature dependence of the resistance of an Al bar realized on a different chip from the same wafer. We measure a critical temperature of $1.25$~K and a normal state resistance of $ 15.5~\Omega$. The kinetic sheet inductance  $L_{\textrm{kin}/\square}$ is then obtained through the low frequency limit of the Mattis-Bardeen screening theory~\cite{Mattis1958,Tinkham2004,Annunziata2010}:
\begin{equation}
  L_{\textrm{kin}/\square} = \frac{\hbar R_{\textrm{n}/\square}}{\pi\Delta_0}\tanh^{-1}\left(\frac{\Delta_0}{2k_BT} \right) \text{.}
  \label{eq:Mattis}
\end{equation}
Here, $R_{\textrm{n}/\square}$ is the normal state sheet resistance, $\Delta_0$ the zero-temperature BCS gap and $T$ the absolut etemperature. Using Eq.~\ref{eq:Mattis} we extract $ L_{\textrm{kin}/\square} \approx 5$nH.

\section{Retrapping versus Switching Current}
\label{sec:Retrapping versus Switching Current}
\begin{figure}[h]
	\centering
	\includegraphics[width=\columnwidth]{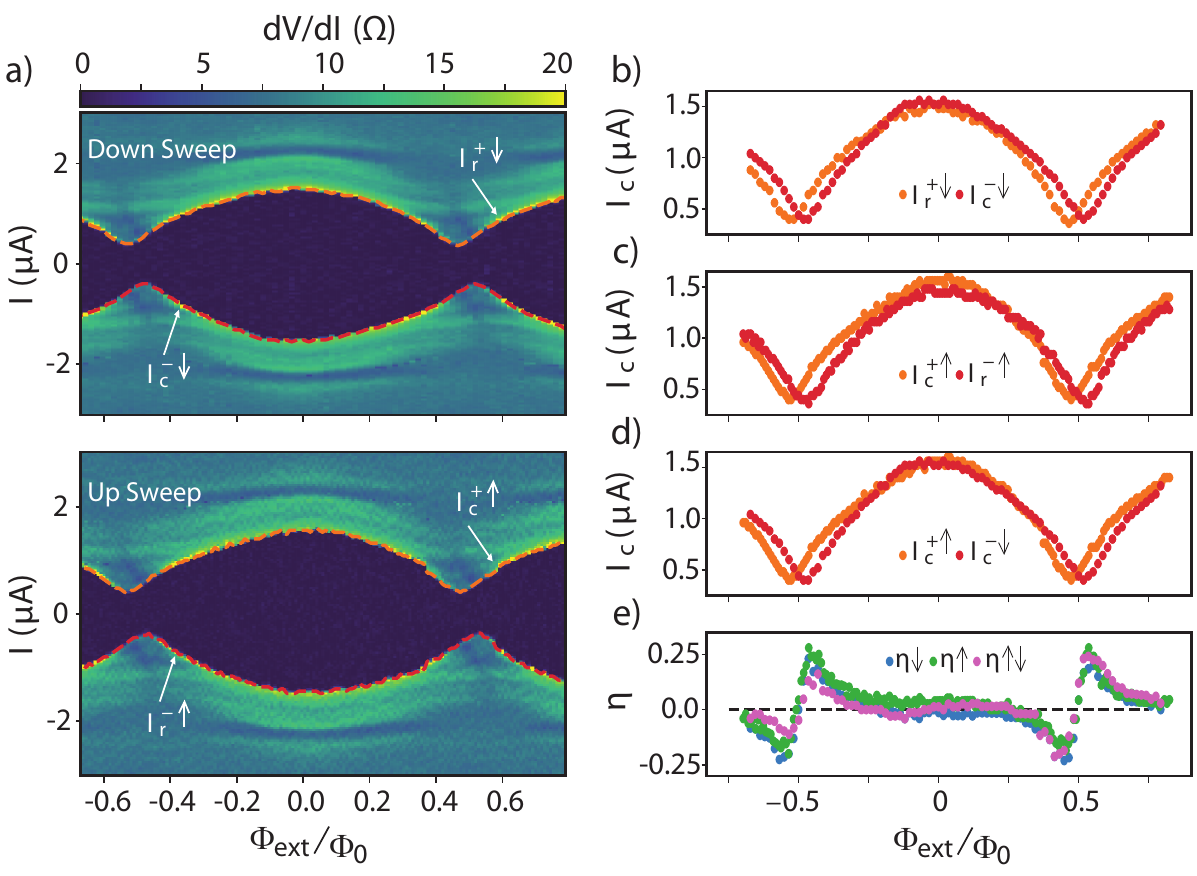}
	\caption{(a) Two differential resistance plots of the SQUID device for the same gate settings as a function of external flux $\Phi_{\textrm{ext}}$ and current bias $I$.
        In the upper plot the current was swept downwards from positive to negative values, while in the lower it was swept upwards. (b) and (c) compare the critical current values $I_{\textrm{c}}$ with the retrapping ones $I_{\textrm{r}}$, obtained from (a) and (b) at the position of the peaks in $dV/dI$. The arrows $\uparrow,\downarrow$ indicate the sweep direction. (d) compares $I_{\textrm{c}}^{+\uparrow}$ with $I_{\textrm{c}}^{-\downarrow}$ and in (e) the diode efficiency is shown for three ways using the data in (b)-(d).
    }
	\label{fig:S3}
\end{figure}

\vspace{-0.3cm}
In Fig.~\ref{fig:S3} we compare the switching current with the retrapping current values. We show that the two values coincide in this experiment to a good accuracy. We think that this is due to the low parallel resistor which keeps the voltage over the junction small in the normal state, hence, reducing overheating effects. Additionally, the shunt resistor adds damping at the plasma frequency of the junctions, which reduces the quality factor.

The two measurements in Fig.~\ref{fig:S3}(a) were obtained for exact the same parameter settings, except for the direction of current-bias sweep. In the upper (lower) measurement the current was decreased (increased) starting with positive (negative) values at $+3$~$\mu$A ($-3$~$\mu$A) and sweeping down (up) to $-3$~$\mu$A ($+3$~$\mu$A). (b) shows the critical and retrapping current, $I_{\textrm{c}}$ and $I_{\textrm{r}}$, extracted from the downsweep data at positions where the differential resistance shows a peak. (c) shows the same, but extracted from the upsweep data. On sweeping downwards, we denote the negative critical current as $I_{\textrm{c}}^{-\downarrow}$ and the positive retrapping current as $I_{\textrm{r}}^{+\downarrow}$. In analogy, on sweeping upwards, the positive critical current is denoted by $I_{\textrm{c}}^{+\uparrow}$ and the negative retrapping current by $I_{\textrm{r}}^{-\uparrow}$. In (d) we compare the positive and negative critical currents, both obtained in a proper way using oppositive sweep directions.

Now we can compare the extracted diode efficiency for three cases: i) for the case when we extract the critical currents from sweeping the current bias into negative direction only, $\eta^\downarrow$, ii) into positive direction only, $\eta^\uparrow$, and iii), when we deduce the critical current properly, $\eta^{\uparrow\downarrow}$. The three curves are directly obtained from the graphs (b)-(d). All three methods yield qualitatively the same efficiencies with no significant differences. Importantly, one clearly cannot say that $\eta^{\uparrow\downarrow}$ would yield in general lower efficiencies.

\section{SQUID Oscillations at Different Gate Voltages}
\label{sec:SQUID Oscillations at Different Gate Voltages}
\begin{figure}[h]
	\centering
	\includegraphics[width=\columnwidth]{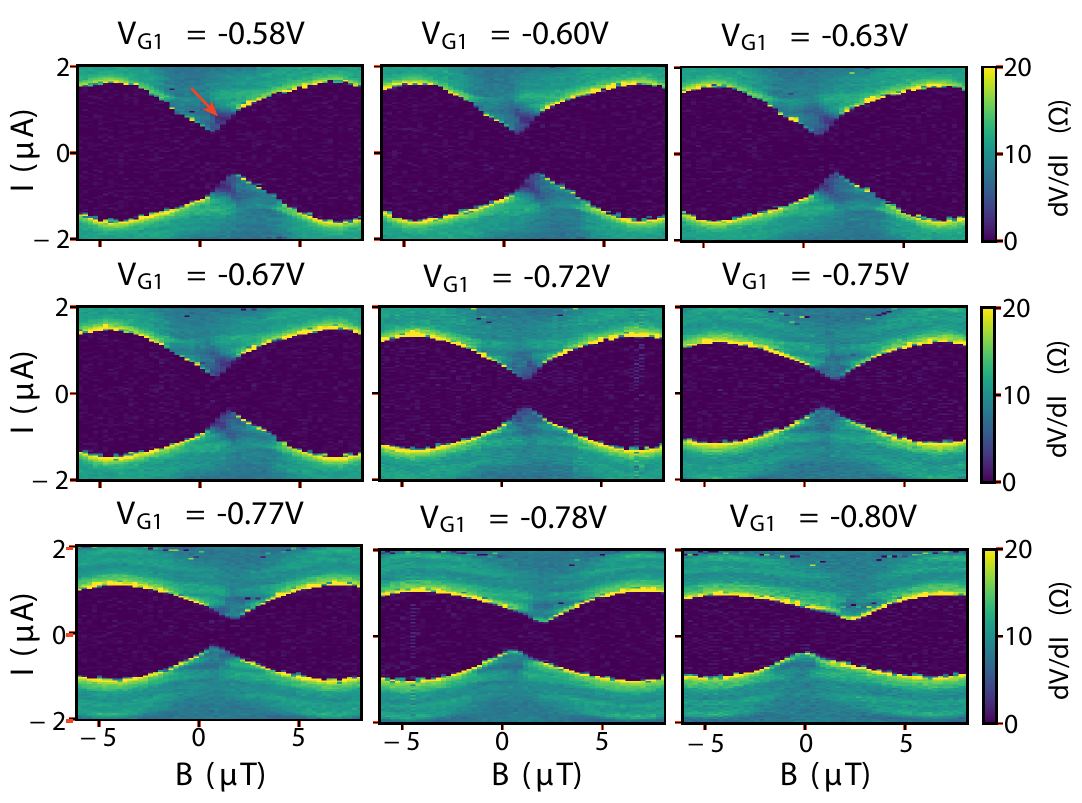}
	\caption{SQUID oscillation at different gate voltage configurations. $V_{\mathrm{G2}}$ is fixed at $-0.5$~V, while $V_{\mathrm{G1}}$ is swept from
       $-0.57$~V to $-0.8$~V. The asymmetry in the SQUID oscillations follows the asymmetry in critical current between the two junctions. We have $I_{\textrm{c1}}(V_{\mathrm{G1}}=-0.57~V) > I_{\textrm{c2}}(V_{\mathrm{G2}}=-0.5~V) $ and
       $ I_{\textrm{c1}}(V_{\mathrm{G1}}=-0.8~V) < I_{\textrm{c2}}(V_{\mathrm{G2}}=-0.5~V) $.}
	\label{fig:S4}
\end{figure}

\vspace{-0.3cm}
In this appendix we show how the SQUID pattern develops when the critical current of one junction is tuned from being larger, equal and finally smaller than the critical current of the other junction. Fig.~\ref{fig:S4} shows the differential resistance of the SQUID as a function of current bias and perpendicular magnetic field. $V_{\mathrm{G2}}$ is fixed at $-0.5$~V, while $V_{\mathrm{G1}}$ is swept from $-0.57$~V to $-0.8$~V. As extracted from Fig.~\ref{fig:Fig1}(c), $I_{\textrm{c2}}(V_{\mathrm{G2}}=-0.5)$ $\sim 720$~nA, while $I_{\textrm{c1}}(V_{\mathrm{G1}}=-0.57)$ $\sim 1.12$~$\mu$A and $I_{\textrm{c1}}(V_{\mathrm{G1}}=-0.8)$ $\sim 360$~nA (gate voltages are given in units of V). \\

The sign of the diode efficiency is mirrored with respect the magnetic field value corresponding to half flux quantum when the critical current asymmetry $ \alpha $ between the two junctions changes sign. We also notice a dip in differential resistance developing around half flux quantum that evolves with $\alpha$ (see arrow in Fig.~\ref{fig:S4}).

\section{Model Including Loop Inductances}
\label{sec:Model Including Loop Inductances}
\vspace{-0.3cm}

As introduced in the main text, we model the current-phase relation of a single junction $\textrm{i} \in [1,2]$ with
\begin{equation}
  I_{\textrm{i}}(\varphi_1)=\frac{N_{\textrm{i}}\tau^*_{\textrm{i}} e \Delta}{\hbar}\frac{\sin(\varphi_{\textrm{i}})}{\sqrt{1-\tau^*_{\textrm{i}} \sin^2(\varphi_{\textrm{i}}/2)}},
\label{eq:eqS1}
\end{equation}
where $N_{\textrm{i}}$ stands for the number of channels and $\tau^*_{\textrm{i}}$ for an effective transmission probability of junction $\mathrm{i}$.
The more general approach would be to assume a distribution function for the transmission probability of each channel. To avoid this complication we assume that all channels have the same transmission probability $\tau^*_i$.

We introduce the normalization parameter $A_{\textrm{i}}$ as
\begin{equation}
  A_{\textrm{i}}:=max_{\varphi_{\textrm{i}}}\left\{\frac{\sin(\varphi_{\textrm{i}})}{\sqrt{1-\tau^*_{\textrm{i}} \sin^2(\varphi_{\textrm{i}})}} \right\}.
\label{eq:eqS2}
\end{equation}
Note, that $A_{\textrm{i}}$ only depends on $\tau^*_{\textrm{i}}$. We thus get the normalized CPR as
\begin{equation}
  I_{\textrm{i}}(\varphi_{\textrm{i}})=\frac{I_{\textrm{ci}}}{A_{\textrm{i}}}\frac{\sin(\varphi_{\textrm{i}})}{\sqrt{1-\tau^*_{\textrm{i}} \sin^2(\varphi_{\textrm{i}}/2)}}.
\label{eq:eqS3}
\end{equation}
In this notation of the CPR, $N$ has been replaced by the critical current $I_{\textrm{c}}$, which appears now explicitly.\\

Flux quantization in the loop imposes:
\begin{equation}
  \varphi_1-\varphi_2 = 2\pi \Phi/\Phi_0.
\label{eq:eqS4}
\end{equation}
Here, the total flux in the loop $\Phi$ is given by the external flux $\Phi_{\mathrm{ext}}$ and the contributions from the screening currents expressed through the loop inductances, $L_1$ and $L_2$, that belong to the two branches. If mutual inductances are considered, too, one has to introduce new effective inductances $L_1^\prime=L_1-M$ and $L_2^\prime=L_2-M$, where $M$ describes the mutual inductance. We obtain for the total flux:
\begin{equation}
  \Phi = \Phi_{\mathrm{ext}} - L_1'I_1(\varphi_1) + L_2'I_2(\varphi_2)
\label{eq:eqS5}
\end{equation}
Therefore, Eq.~\ref{eq:eqS4} now reads:
\begin{equation}
  \varphi_1-\varphi_2 = \varphi_{\mathrm{ext}} + \frac{2\pi}{\Phi_0}\left(L_2'I_2(\varphi_2)-L_1'I_1(\varphi_1)\right).
\label{eq:eqS6}
\end{equation}

Our simulations show, however, that the effect of the mutual inductance can be neglected in our experiment. Hence, there are six remaining parameters in the problem: $I_{\textrm{c1}}$, $I_{\textrm{c2}}$, $\tau^*_1$, $\tau^*_2$, $L_1$, and $L_2$. Since the appearance of the SDE in a SQUID is related to asymmetries, we introduce three asymmetry parameters:
\begin{equation}
  \alpha:=\frac{I_{\textrm{c1}}-I_{\textrm{c2}}}{I_{\textrm{c1}}+I_{\textrm{c2}}},
\label{eq:eqS7}
\end{equation}
\begin{equation}
  \beta:=\frac{\tau^*_1-\tau^*_2}{\tau^*_1 + \tau^*_2},
\label{eq:eqS8}
\end{equation}
\begin{equation}
  \gamma:=\frac{L_1-L_2}{L_1+L_2}.
\label{eq:eqS9}
\end{equation}
The new set of parameters is now given by the three asymmetries and the average values of the two junctions for the critical current $\bar{I_{\textrm{c}}}$, the transmission probability $\bar{\tau}$ and the inductance $\bar{L}$.\\

To find the critical current one has to find the maximum or minimum of the total super\-current:
\begin{equation}
  I(\varphi_1,\varphi_2)=I_1(\varphi_1)+I_2(\varphi_2).
\label{eq:eqS10}
\end{equation}
Making use of Eq.~\ref{eq:eqS6}, we get:
\begin{eqnarray}
  I(\varphi_1,I) & = & I_1(\varphi_1)+I_2(\varphi_1-\varphi_{\mathrm{ext}} +\kappa L_1I_1(\varphi_1) \nonumber \\
                 &   & -\kappa L_2(I-I_1(\varphi_1))),
\label{eq:eqS11}
\end{eqnarray}
with $\kappa = 2\pi/\Phi_0$. In the latter form, we have eliminated $\varphi_2$ using the fluxoid condition. However, due to the loop inductances, the equation for the total current $I$ is now itself implicitly dependent on $I$. One can still solve this equation recursively or by introducing Lagrange multipliers to then search for the maximum or minimum currents, yielding $I_{\textrm{c}}^+$ and $I_{\textrm{c}}^-$ \cite{Tesche1977}.\\

To find $I_{\textrm{c}}^+$ numerically, we preset the value of $I$, $0\le I \le 2\bar{I_{\textrm{c}}}$, starting with a small one and search for solutions $\varphi_1$ of Eq.~\ref{eq:eqS11}. If solutions exist, we increment $I$ by a small step $\delta I$ until there are no solutions $\varphi_1$ anymore. This defines $I_{\textrm{c}}^+$. In analogy we obtain $I_{\textrm{c}}^-$.\\

\section{Comparison to Diode Effect due to Loop Inductances}
\label{sec:Comparison to Diode Effect by Loop Inductance}
\begin{figure}[h]
	\centering
	\includegraphics[width=\columnwidth]{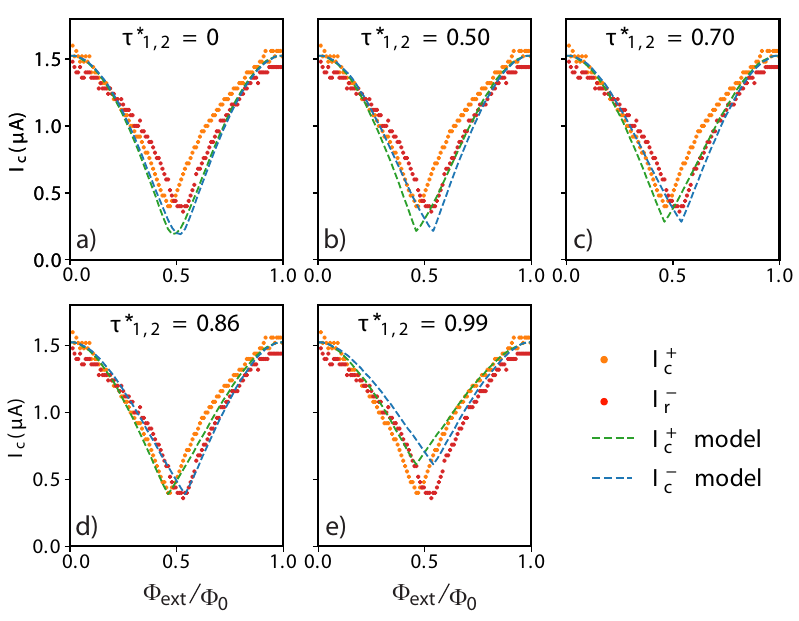}
	\caption{Sequence of simulations for $I_{\textrm{c}}^+$ (green dashed curves) and $I_{\textrm{c}}^-$ (blue dashed curves) to measured (upsweep)
       data $I_{\textrm{c}}^+$ (orange dots) and $I_{\textrm{r}}^-$ (red dots). In (a) a sinusoidal CPR with the estimated loop inductance asymmetry is considered, while in graphs (b)-(e) the effective  transparencies $\tau^*$ of the junctions ar increased. Further details are given in the text.}
	\label{fig:S5}
\end{figure}

\vspace{-0.3cm}
Here, we present a comparison of the measured critical currents $I_{\textrm{c}}^+$ and $I_{\textrm{c}}^-$  shown in Fig.~\ref{fig:Fig2}(c) with model simulations. Specifically, we discuss the effect of the loop inductance and its asymmetry on the SDE. The comparison shows that the SDE can poorly be reproduced taking only the loop inductances into account. This is shown in figure Fig.~\ref{fig:S5}.\\

Figure~\ref{fig:S5} shows a sequence of simulations, blue and green dashed curves, to a set of measurements of $I_{\textrm{c}}^+$ (orange) and $I_r^-$ (red). In all five simulations the
critical currents $I_{\textrm{c1}}$ and $I_{\textrm{c2}}$ of the two junctions are taken from the experiment, from Fig.~\ref{fig:Fig1}c. Since $V_{\mathrm{G1}}=V_{\mathrm{G2}}=0$ we obtain $I_{\textrm{c1}}=0.87$~$\mu$A and $I_{\textrm{c2}}=0.67$~$\mu$A. In (a) we assume sinusoidal CPRs for both junctions JJ$_1$ and JJ$_2$, and we take the simulated loop inductances into account. Due to the slight asymmetry in loop inductance a small SDE appears. However, this effect is far smaller than what has been measured. Hence, one cannot fit the measurement with the loop inductance asymmetry alone. In (b)-(e) we keep the loop inductances as estimated, but change to non-sinusoidal CPRs by increasing $\tau^*_1=\tau^*_2$ to appreciable values ranging from $0.5-0.99$, indicated in the figures. As before, we obtain the blue and green dashed curves taking the known critical currents $I_{\textrm{c1}}$ and $I_{\textrm{c2}}$ of the two junctions. The best match in this sequence is found for $\tau^*_1=\tau^*_2\approx 0.86$. One can see that the model matches the key features of the experiment very well. However, there are deviations, as seen by the stronger curvature that the measurement points display as compared to the model. The model assumes an almost triangular shape for very large transparencies $\tau^*_1=\tau^*_2\approx 0.99$
 These differences are yet not understood

\section{Conditions for a Diode Effect in a SQUID Device}
\label{sec:Conditions for a Diode Effect in a SQUID Device}
\begin{figure}[h]
	\centering
	\includegraphics[width=\columnwidth]{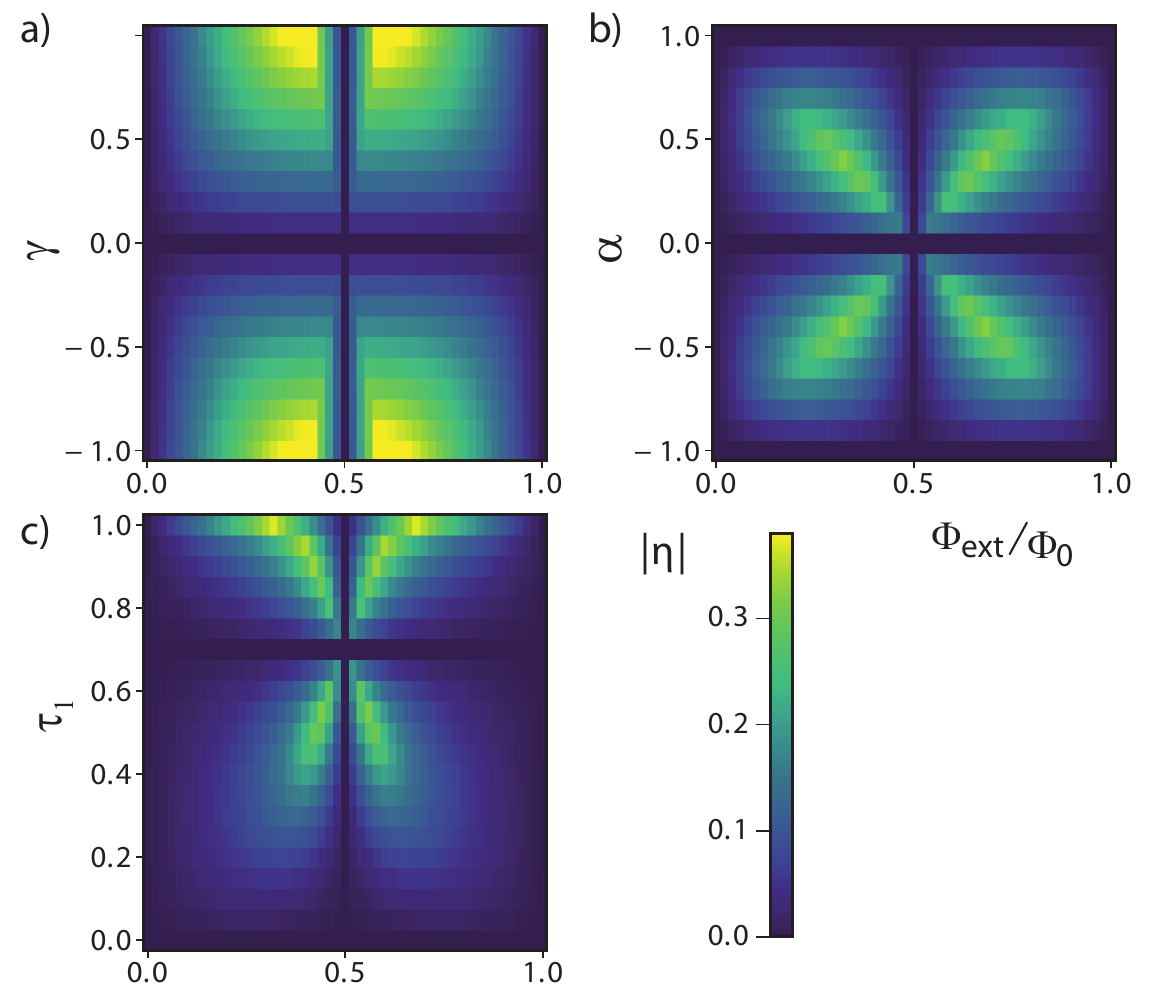}
	\caption{Magnitude of the diode efficiency $|\eta|$ as a function of the applied external flux $\Phi_{\textrm{ext}}$ expressed in number of magnetic flux quanta
       $\Phi_0$, numerically calculated for a SQUID with two sinusoidal CPRs with an asymmetry (a) in loop inductance $\gamma$ and (b) in critical current $\alpha$. The inductances were chosen such that $\varphi_{\textrm{L}} = \pi$. In (c), $|\eta|$ is plotted for a SQUID without loop inductances and two JJs, each with a non-sinusoidal single-channel CPR, as a function of $\tau_1$ and normalized external flux for $\tau_2 = 0.7$ and for $\alpha=\gamma=\varphi_{\textrm{L}}=0$.}
	\label{fig:S6}
\end{figure}

\vspace{-0.3cm}
The following three figures illustrate that an asymmetry is required to obtain a SDE. in Fig.~\ref{fig:S6}(a) and (b) sinusoidal CPRs are assumed. In (a) the loop inductance asymmetry $\gamma$ is varied, while the critical-current asymmetry $\alpha = 0$. In contrast, in (b) $\alpha$ is varied, while $\gamma=0$. The loop inductance has been chosen such that the average phase drop over the inductor $\varphi_{\mathrm{L}}=4\pi\bar{I_{\textrm{c}}}\bar{L}\Phi_0$ assumes a large value of $\varphi_{\mathrm{L}}=\pi$. In (c) a SQUID with two single-channel non-sinusoidal CPRs with different transmission probabilities $\tau_{1,2} \not = 0$ (asymmetry $\beta \not = 0$) are considered, while $\alpha=\gamma=\varphi_{\mathrm{L}}=0$.

In general, it is seen that the diode efficiency is zero at the symmetry points corresponding in (a) to $\gamma =0$, in (b) to $\alpha =0$  and in (c) to $\tau_1 = \tau_2$. Further on, $\eta = 0$ for $\varphi_{\mathrm{ext}}=2\pi\Phi_{\textrm{ext}}/\Phi_0 = 0$, $\pi$, and $2\pi$. For these cases one can show that the CPR of the SQUID is odd in the phase difference $\varphi$. This follows from Eq.~\ref{eq:ISquid2} and the fact that $I_{\textrm{1}}(\varphi)$ and $I_{\textrm{2}}(\varphi)$ are odd functions in $\varphi$.
In addition, we note that the position of maximum diode efficiency in flux depends on what kind of asymmetry dominates. It can take up values $> 30$~$\%$.

To obtain a SDE in a SQUID loop, an asymmetry is required. This we have illustrated in the previous figure Fig.~\ref{fig:S6} where out of the three asymmetry parameters $\alpha$, $\beta$, $\gamma$ only one was different from zero. In the following table we show under which conditions the SDE appears depending on all three asymmetry parameters. The table shows that at least one symmetry has to be broken to get the SDE effect. This is a sufficient condition for almost all cases. There is only one exception. It arises for sinusoidal CPRs where a difference in critical currents of the two junctions is not enough for a SDE to appear.\\

\begin{table}[H]
  \centering
	\caption{\label{table2} Conditions for obtaining a super\-conducting diode-effect (SDE). In the first column $\tau^*=0$ is used to refer to a sinusoidal CPR, while
             $\tau^*\not=0$ indicates a highly transmissive CPR containing higher order terms in the CPR. If $\bar{L}=0$, loop inductances are not  considered, while they play a role in the entries where $\bar{L}\not=0$. $\alpha$ ($\beta$) denotes the asymmetry in $I_{\textrm{c}}$ ($\tau^*$) of the two JJs, while $\gamma$ denotes the asymmetry in the loop inductances in the two arms of the SQUID.\\}
	\begin{minipage}{0.4\textwidth}
		\begin{ruledtabular}
			\begin{tabular}{|c|c|c|c|c|c|}
				$\tau^*$ & $\beta$ & $\alpha$ & $\bar{L}$ & $\gamma$ & SDE \\
				\hline
				$0$ & n.a. & $0$ & $0$ & n.a. & no \\
				\hline
				$0$ & n.a. & $0$ & $\neq 0$ & $0$ & no \\
				\hline
				$0$ & n.a. & $0$ & $\neq 0$ & $\neq 0$ & yes \\
				\hline
				$0$ & n.a. & $\neq 0$ & $0$ & n.a. & no \\
				\hline
				$0$ & n.a. & $\neq 0$ & $\neq 0$ & $0$ & yes \\
				\hline
				$0$ & n.a. & $\neq 0$ & $\neq 0$ & $\neq 0$ & yes \\
				\hline
				$\neq 0$ & $0$ & $0$ & $0$ & n.a. & no \\ 
				\hline
				$\neq 0$ & $0$ & $0$ & $\neq 0$ & $0$ & no \\
				\hline
				$\neq 0$ & $0$ & $0$ & $\neq 0$ & $\neq 0$ & yes \\
				\hline
				$\neq 0$ & $0$ & $\neq 0$ & $0$ & n.a. & yes \\
				\hline
				$\neq 0$ & $0$ & $\neq 0$ & $\neq 0$ & $0$ & yes \\
				\hline
				$\neq 0$ & $0$ & $\neq 0$ & $\neq 0$ & $\neq 0$ & yes \\
				\hline
				$\neq 0$ & $\neq 0$ & $0$ & $0$ & n.a. & yes \\ 
				\hline
				$\neq 0$ & $\neq 0$ & $0$ & $\neq 0$ & $0$ & yes \\
				\hline
				$\neq 0$ & $\neq 0$ & $0$ & $\neq 0$ & $\neq 0$ & yes \\
				\hline
				$\neq 0$ & $\neq 0$ & $\neq 0$ & $0$ & n.a. & yes \\
				\hline
				$\neq 0$ & $\neq 0$ & $\neq 0$ & $\neq 0$ & $0$ & yes \\
				\hline
				$\neq 0$ & $\neq 0$ & $\neq 0$ & $\neq 0$ & $\neq 0$ & yes
			\end{tabular}
		\end{ruledtabular}
	\end{minipage}
\end{table}
\vspace{-0.3cm}


%

\end{document}